\documentclass[12pt]{article}

\usepackage{amsmath}
\usepackage{amssymb}
\usepackage[mathcal]{euscript}
\usepackage{graphicx,psfrag}
\usepackage{epsfig}

\usepackage{graphicx}
\usepackage{color}
\usepackage{epic,eepic}

\usepackage{graphicx}
\usepackage{color}

\usepackage{amsmath}
\usepackage{latexsym}

\newtheorem{theorem}{Theorem}

\usepackage{graphicx}

\begin{document}

\title{Classification of 3-dimensional integrable scalar discrete 
equations\footnote{Supported by the DFG Research Unit 565 ``Polyhedral Surfaces'' (TU-Berlin)}}

\author{
  S.P.~Tsarev\thanks{On leave from: Krasnoyarsk State Pedagogical
  University, Russia. SPT acknowledges partial financial support from
the grant of Siberian Federal University (NM-project $N^o$~45.2007) and 
the RFBR grant   06-01-00814.}
\   and T.~Wolf}

\maketitle
\begin{center}
Department of Mathematics \\
Technische Universit\"at Berlin \\
Berlin, Germany \\[1ex]
and \\[1ex]
Department of Mathematics, Brock University \\
500 Glenridge Avenue, St.Catharines, \\
Ontario, Canada L2S 3A1\\
e-mails: \\%[1ex]
\texttt{tsarev@math.tu-berlin.de} \ \ \ \  %\\
\texttt{sptsarev@mail.ru}\\
\texttt{twolf@brocku.ca}\\
\end{center}

\medskip

\begin{abstract}

    We classify all integrable 3-dimensional scalar discrete
quasilinear equations $Q_3=0$ on an elementary cubic cell of the
lattice ${\mathbb Z}^3$. An equation $Q_3=0$ %of such form
is called integrable if it may be consistently imposed on all
$3$-dimensional elementary faces of the lattice ${\mathbb Z}^4$.
Under the natural requirement of invariance of the equation under the
action of the complete group of symmetries of the cube we prove that
the only nontrivial (non-linearizable) integrable equation from this
class is the well-known dBKP-system.


                   {\sc Reduce}, {\sc Form}, {\sc Crack}.
\end{abstract}

\section{Introduction}

Although definitively shaped a decade ago, discrete differential
geometry (see e.g.\ \cite{BS,ABOP}) has already provided much
insight into structures that are fundamental both to classical
differential geometry and to the theory of integrable PDEs. In
addition to such purely mathematical fields, results in discrete
differential geometry have a great potential in computer graphics
and architectural design: it turns out that discrete surfaces
parameterized by discrete conjugate lines and discrete curvature
lines --- the basic structures in discrete differential geometry ---
have superior approximation properties and other useful features
(see \cite{Pot}).

\begin{figure}[http]
\begin{center}
\begin{picture}(0,0)%
\includegraphics{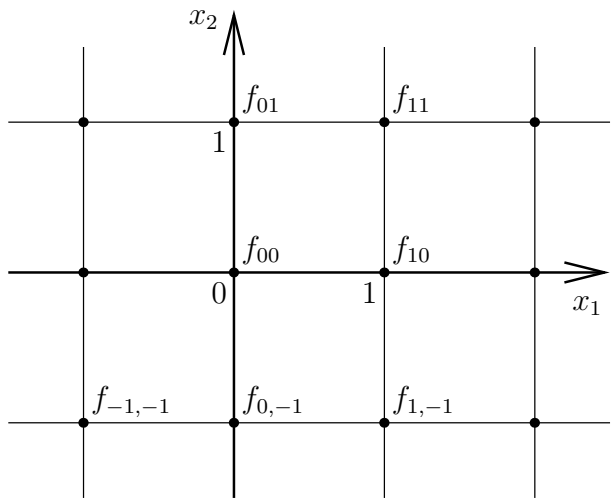}%
\end{picture}%
\setlength{\unitlength}{4144sp}%
\begingroup\makeatletter\ifx\SetFigFont\undefined%
\gdef\SetFigFont#1#2#3#4#5{%
  \reset@font\fontsize{#1}{#2pt}%
  \fontfamily{#3}\fontseries{#4}\fontshape{#5}%
  \selectfont}%
\fi\endgroup%
\begin{picture}(3858,2977)(4479,-5033)
\put(5716,-3841){\makebox(0,0)[lb]{\smash{{\SetFigFont{12}{14.4}{\familydefault}{\mddefault}{\updefault}{\color[RGB]{0,0,0}$0$}%
}}}}
\put(5896,-3571){\makebox(0,0)[lb]{\smash{{\SetFigFont{12}{14.4}{\familydefault}{\mddefault}{\updefault}{\color[RGB]{0,0,0}$f_{00}$}%
}}}}
\put(5896,-2671){\makebox(0,0)[lb]{\smash{{\SetFigFont{12}{14.4}{\familydefault}{\mddefault}{\updefault}{\color[RGB]{0,0,0}$f_{01}$}%
}}}}
\put(6796,-3571){\makebox(0,0)[lb]{\smash{{\SetFigFont{12}{14.4}{\familydefault}{\mddefault}{\updefault}{\color[RGB]{0,0,0}$f_{10}$}%
}}}}
\put(6796,-4471){\makebox(0,0)[lb]{\smash{{\SetFigFont{12}{14.4}{\familydefault}{\mddefault}{\updefault}{\color[RGB]{0,0,0}$f_{1,-1}$}%
}}}}
\put(5896,-4471){\makebox(0,0)[lb]{\smash{{\SetFigFont{12}{14.4}{\familydefault}{\mddefault}{\updefault}{\color[RGB]{0,0,0}$f_{0,-1}$}%
}}}}
\put(4996,-4471){\makebox(0,0)[lb]{\smash{{\SetFigFont{12}{14.4}{\familydefault}{\mddefault}{\updefault}{\color[RGB]{0,0,0}$f_{-1,-1}$}%
}}}}
\put(6616,-3841){\makebox(0,0)[lb]{\smash{{\SetFigFont{12}{14.4}{\familydefault}{\mddefault}{\updefault}{\color[RGB]{0,0,0}$1$}%
}}}}
\put(5716,-2941){\makebox(0,0)[lb]{\smash{{\SetFigFont{12}{14.4}{\familydefault}{\mddefault}{\updefault}{\color[RGB]{0,0,0}$1$}%
}}}}
\put(7876,-3886){\makebox(0,0)[lb]{\smash{{\SetFigFont{12}{14.4}{\familydefault}{\mddefault}{\updefault}{\color[RGB]{0,0,0}$x_1$}%
}}}}
\put(6796,-2671){\makebox(0,0)[lb]{\smash{{\SetFigFont{12}{14.4}{\familydefault}{\mddefault}{\updefault}{\color[RGB]{0,0,0}$f_{11}$}%
}}}}
\put(5581,-2176){\makebox(0,0)[lb]{\smash{{\SetFigFont{12}{14.4}{\familydefault}{\mddefault}{\updefault}{\color[RGB]{0,0,0}$x_2$}%
}}}}
\end{picture}%
\end{center}
\caption{${\mathbb Z}^2$ lattice} \label{fig1}
\end{figure}

In this paper we consider the cubic lattice ${\mathbb Z}^n$ with
vertices at integer points
in the $n$-dimensional space ${\mathbb R}^n=\{(x_1,\ldots, x_n)| x_s \in {\mathbb R}\} \}$.
With each vertex (with integer coordinates $(i_1, \ldots , i_n)$, $i_s \in {\mathbb Z}$)
we associate a scalar field variable $f_{i_1 \ldots i_n}\in {\mathbb C}$.

In what follows we need to consider the elementary cubic cell $K_n =
\left\{ (i_1, \ldots, i_n)\right. | $ $\left. i_s \in \{0,1\} \right\}$ of the lattice
${\mathbb Z}^n$. The field variables $f_{i_1 \ldots i_n}$
%$i_s \in \{0,1\}$ 
are associated to its $2^n$ vertices.  We will use the short
notation ${\mathbf{f}}$ for the set $(f_{{00\ldots 0}}, \ldots , f_{11
\ldots 1})$ of all these $2^n$ variables.

An $n$-dimensional discrete system of the type considered here is given by
an equation of the form
\begin{equation}\label{eq0}
 Q_n({\mathbf{f}}) = 0,
\end{equation}
on the field variables on the elementary cubic cell $K_n$. For the
other elementary cubic cells of ${\mathbb Z}^n$ the equation is the
same, after shifting the indices of ${\mathbf{f}}$ suitably (see
Figure~\ref{fig1}).

In the last two decades the study of special classes of (\ref{eq0})
which are ``integrable'' (in one sense or another) has become very
popular. We give below only a brief account of the current state of
this field of research, for a more detailed account cf.
\cite{ABS}--\cite{ABOP} and the references given therein.  In fact,
discrete integrable systems underlie many classical integrable
nonlinear PDEs, like the Krichever-Novikov equation and other examples,
the latter appear as a continuous limit along some of the discrete
directions. Other well known classes of integrable geometric objects
(with $n=3$), like minimal surfaces, conjugate nets, constant
curvature surfaces, Moutard nets, isothermic surfaces, orthogonal
curvilinear coordinates etc., are also obtained as some smooth limits
along any two of the three directions of the respective
discrete system. The remaining third discrete direction automatically
provides us with a transformation known in the classical continuous
geometric context as Jonas/Ribaucour/B\"acklund transformation between
surfaces of the given class (see \cite{ABS}--\cite{ABOP} for more
details). On the other hand, starting from the classical theorems on
non-linear superposition principles and permutability of the
aforementioned transformations between smooth surfaces of one of these
types we obtain precisely the underlying discrete system.  One of the
cornerstones of the discrete differential geometry (the idea to look
for cubic nonlinear superposition formulas of B\"acklund
transformations of nonlinear integrable PDEs) was laid down in
\cite{GT}. The duality between the smooth objects in any of the
geometric classes of integrable smooth surfaces mentioned above and
their B\"acklund-type transformations is therefore put into a
symmetric form of a single discrete $n$-dimensional system and is
encoded as the notion of \textit{$(n+1)$-dimensional consistency}
\cite{BS}:

\textit{An $n$-dimensional discrete equation (\ref{eq0}) is called
consistent, if it may be imposed in a consistent way on all
$n$-dimensional faces of a $(n+1)$-dimensional cube}.

This can be also understood as the possibility to take ${\mathbb
Z}^{n+1}$ and prescribe the $n$-dimensional equation (\ref{eq0}) to
hold on every $n$-dimensional face of every elementary
$(n+1)$-dimensional cube (of size 1, with edges parallel to the
coordinate axes) \textit{without} side relations to appear.  For this
reason (\ref{eq0}) is often called a ``face formula''.  A precise
definition of consistency, suitable for the class of discrete
equations treated in this paper, will be formulated in the next
section.

This paper is devoted to application of computer algebra systems {\sc
Reduce} and {\sc Form} (\cite{FORM}), in particular the {\sc Reduce} package
{\sc Crack} (\cite{CrackOver,CrackOnline}), to the classification of
3-dimensional integrable discrete systems.

The paper is organized as follows. In Section~\ref{sec2} we give a
brief description of the known results on 2-dimensional integrable
scalar discrete equations of type (\ref{eq0}) and the precise
definition of $(n+1)$-dimensional consistency condition for such
discrete $n$-dimensional systems.

Section~\ref{sec3} is devoted to the classification of symmetry types
of quasilinear equations (\ref{eq0}) for dimensions $n=2,3,4$.

In Section~\ref{sec4} we describe the results of our computations
(Theorem~\ref{th-nonex}):
the only nontrivial (non-linearizable) integrable
scalar quasilinear $3d$-face equation invariant w.r.t.\ the complete
group of symmetries of the cube is given by the formula (\ref{Q111})
below.

Appendices~A--E describe the technical details of the computations.

\section{The setup}\label{sec2}

The simplest but very important class of 2-dimensional integrable face formulas
was investigated in detail in \cite{ABS,ABS2}. They have the form
\begin{equation}\label{face2}
 Q(f_{00},f_{10},f_{01},f_{11}) =0,
\end{equation}
where $f_{ij}$ are scalar fields attached to the vertices of a
square (see Fig.~\ref{figur1}) with two main requirements:

\textbf{1) Quasilinearity.} (\ref{face2}) is affine linear w.r.t.\
every  $f_{ij}$,  i.e. $Q$ has degree 1 in any of its four
variables: $Q= c_1(f_{10},f_{01},f_{11})f_{00} +
c_2(f_{10},f_{01},f_{11})= c_3(f_{00},f_{01},f_{11})f_{10} +
c_4(f_{00},f_{01},f_{11})=\ldots = q_{1111}f_{00}f_{10}f_{01}f_{11} + q_{1110}f_{00}f_{10}f_{01}
+ q_{1101}f_{00}f_{10}f_{11} + \ldots + q_{0000}$.

\textbf{2) Symmetry.} Equation (\ref{face2}) should be invariant
w.r.t.\ the symmetry group of the square or its suitably chosen
subgroup.

%-----------------------------------------------------------------
\begin{center}
\begin{figure}[http]
\setlength{\unitlength}{0.025em}
   \begin{minipage}[t]{150pt}
\begin{center}
\begin{picture}(200,220)(0,0)
 \put(0,0){\circle*{20}}    \put(150,0){\circle*{20}}
 \put(0,150){\circle*{20}}  \put(150,150){\circle*{20}}
 \path(0,0)(150,0)       \path(0,0)(0,150)
 \path(150,0)(150,150)   \path(0,150)(150,150)
 \put(-40,-30){$f_{00}$}
 \put(-40,170){$f_{01}$}
 \put(155,-30){$f_{10}$}
 \put(155,170){$f_{11}$}
\end{picture}
\end{center}
\caption{Square $K_2$.}\label{figur1}
    \end{minipage}%\hfill
\ \ \ \
\begin{minipage}[t]{200pt}
\begin{center}
\begin{picture}(200,220)(0,0)
 \put(0,0){\circle*{20}}    \put(150,0){\circle*{20}}
 \put(0,150){\circle*{20}}  \put(150,150){\circle*{20}}
 \put(50,50){\circle*{20}} \put(50,200){\circle*{20}}
 \put(200,50){\circle*{20}}
 \put(200,200){\circle*{20}}
 \path(0,0)(150,0)       \path(0,0)(0,150)
 \path(150,0)(150,150)   \path(0,150)(150,150)
 \path(0,150)(50,200)    \path(150,150)(200,200)
 \path(50,200)(200,200)
 \path(200,200)(200,50) \path(200,50)(150,0)
 \dashline[+30]{10}(0,0)(50,50)
 \dashline[+30]{10}(50,50)(50,200)
 \dashline[+30]{10}(50,50)(200,50)
 \put(-50,-30){$f_{000}$}
 \put(-50,170){$f_{001}$}
 \put(23,75){$f_{010}$}
 \put(22,220){$f_{011}$}
 \put(160,-25){$f_{100}$}
 \put(160,140){$f_{101}$}
 \put(215,50){$f_{110}$}
 \put(215,220){$f_{111}$}
\end{picture}
\caption{Cube $K_3$.}\label{fig2}
\end{center}
   \end{minipage}
\end{figure}

\end{center}
%-----------------------------------------------------------------

A few other requirements were given in \cite{ABS}, in particular the
formula (\ref{face2}) involved \textit{parameters} attached to the
edges of the square.

The second requirement of symmetry is obviously very important for
the formulation of the condition of $3$-dimensional consistency of (\ref{face2}).
Namely, suppose we have an elementary cube ($3d$-cell of  ${\mathbb Z}^3$, cf.~Fig.~\ref{fig2})
and impose (\ref{face2}) to hold on three ``initial $2d$-faces''
$\{x_1=0\}$: $Q(f_{000},f_{010},f_{001},f_{011}) =0$;
$\{x_2=0\}$: $Q(f_{000},f_{100},f_{001},f_{101}) =0$;
$\{x_3=0\}$: $Q(f_{000},f_{100},f_{010},f_{110}) =0$
(these are used to find $f_{011}$, $f_{101}$, $f_{110}$ from
$f_{000}$, $f_{100}$, $f_{010}$, $f_{001}$).
Then we impose (\ref{face2}) to hold on the other three ``final $2d$-faces''
$\{x_1=1\}$: $Q(f_{100},f_{110},f_{101},f_{111}) =0$;
$\{x_2=1\}$: $Q(f_{010},f_{110},f_{011},f_{111}) =0$;
$\{x_3=1\}$: $Q(f_{001},f_{101},f_{011},f_{111}) =0$;
so for the last field variable $f_{111}$  we can find 3 (apriori)
different rational expressions in terms of the ``initial
data''  $f_{000}$, $f_{100}$, $f_{010}$, $f_{001}$.
The \textit{$3d$-consistency} is the requirement
that these three expressions of  $f_{111}$ in terms of
the initial data should be identically equal. The subtle
point of this process consists in the non-uniqueness of
the mappings of a given square (Fig.~\ref{figur1}) onto
the six $2d$-faces of the cube. The requirement of symmetry
given above guarantees that we can choose any
identification of the vertices of the 2-dimensional faces of the
3-dimensional elementary cube (Fig.~\ref{fig2}) with the vertices of the
``standard'' square where (\ref{face2}) is given;
certainly this identification should preserve the combinatorial
structure of the square (neighbouring vertices remain neighbouring).
In \cite{ABS} a complete classification of $3d$-consistent $2d$-face
formulas (in a slightly different setting) was obtained;
in \cite{ABS2} a similar classification was given for the
case when one does \textit{not} assume that the formula
(\ref{face2}) is the same on all the 6 faces of the
3-dimensional cube.

In the next sections we give a symmetry classification of
all possible $3d$-face formulas defined
on some ``standard'' 3-dimensional cube:
\begin{equation}\label{face3}
 Q(f_{000},f_{100},f_{010},f_{001},f_{110},f_{101},f_{011},f_{111}) =0
\end{equation}
with respect to the complete symmetry group of the cube.
Here, as everywhere in the paper, indices of the field variables
$f_{ijk}$ give the coordinates of the corresponding vertices of
the standard $3d$-cube where our formula (\ref{face3}) is defined.

The requirement of consistency is now formulated similarly to the
$2d$-case: given a $4d$-cube with field values $f_{ijkl}$, $i,j,k,l
\in \{0,1\}$, one should impose the formula (\ref{face3}) on every
$3d$-face of it, by fixing one of the indices $i,j,k,l$, and making
it 0 for the faces which we will call below ``initial faces'', or
respectively 1 for the faces which we will call ``final faces''. One
also needs to fix some mapping from the initial ``standard'' cube
(with the vertices labelled $f_{ijk}$) onto every one of the eight
$3d$-faces (for example $\{f_{i1kl}\}$ on $\{x_2=1\}$). This can
certainly be done using the trivial lexicographic correspondence of
the type $f_{ijk} \mapsto f_{i1jk}$. Geometrically this
lexicographic correspondence is less natural since it is not
invariant w.r.t.\ the symmetry group of a $4d$-cube.  On the other
hand there is an important example of such a non-symmetric formula
corresponding to the discrete BKP equation (\cite{ABS},
equation~(76)). Another possibility to avoid this problem is to
impose the requirement of symmetry. More precisely, if one applies
any one of the transformations from the group of symmetries of the
$3d$-cube, (\ref{face3}) shall be transformed into an equation with
the left hand side \textit{proportional} to the original expression
$Q$: $Q \mapsto \lambda \cdot Q$. Since this symmetry group is
generated by reflections, one has $\lambda^2=1$, so this
proportionality multiplier $\lambda$ may be either $(+1)$ or $(-1)$
for any particular transformation in the complete symmetry group.

From results in \cite{ABS} we know that there are important
$4d$-consistent $3d$-face formulas which are preserved
under a suitable \textit{subgroup} of the complete symmetry group of
the $3d$-cube. No classifiaction of $3d$-face formulas
with such restricted symmetry property
has beet caried out yet.

\section{Symmetry classification}\label{sec3}

Every $n$-dimensional face formula $Q_{n}=0$ which satisfies the
requirement of quasilinearity  has a left hand side of the form
\begin{equation}\label{gf}
 Q_{n}=\sum_{{\cal D}} q_{{\cal D}} \prod_{i_s = 0,1}
    \big(f_{i_1\ldots i_n}\big)^{D_{i_1\ldots i_n}}
\end{equation}
with constant coefficients $q_{{\cal D}}$, where the summation is
taken over all $2^{2^n}$ many  $2^{n}$-tuples ${\cal D}=(D_{00\ldots 0},\ldots ,
D_{11\ldots1})$, each power $D_{i_1\ldots i_n}$ of the respective
vertex variable $f_{i_1\ldots i_n}$ being either 0 or 1.
In other words: the $2^n$ indices of $q_{{\cal D}}$ are the exponents of the $2^n$
vertex field variables $f_{i_1\ldots i_n}$, each exponent $D_{i_1\ldots i_n}$
being $0$ or $1$.
For example,
$Q_2= q_{1111}f_{00}f_{10}f_{01}f_{11} + q_{1110}f_{00}f_{10}f_{01}
+ q_{1101}f_{00}f_{10}f_{11} + \ldots + q_{0000}$ has $2^{2^2} = 16$
terms, $Q_3$ has respectively $2^{2^3} = 256$ terms, and $Q_4$ has
already $2^{2^4} = 65536$ terms.

In this Section we classify $n$-dimensional quasilinear equations
$Q_{n}=0$ for $n=2,3,4$  that are invariant w.r.t. the complete
symmetry group of the respective $n$-dimensional cube.
This problem can be reduced to the enumeration of irreducible
representations of this group in the space of polynomials of the form
(\ref{gf}). Here this is done
in a straightforward way: the group in question is generated by one
reflection w.r.t\ the plane $x_1 = 1/2$  and
$(n-1)$ diagonal reflections w.r.t. the planes $x_1 = x_{s} $,
$s=2, \ldots , n$  (here $x_k$  denote the
coordinates in  ${\mathbb R}^{n}$).

To every reflection $R$ from this generating set we assign $(-)$ or
$(+)$ and require the equality $Q({\mathbf{f}})  = - Q(R({\mathbf{f}}))$
(respectively  $Q({\mathbf{f}})  =  Q(R({\mathbf{f}}))$) to hold
identically in all vertex variables  ${\mathbf{f}}$; this
gives us a set of equations for the coefficients  $q_{{\cal D}}$.
Running through all possible choices of the signs for the generating
reflections we solve the united sets of simple linear equations for
the coefficients   $q_{{\cal D}}$ for every such choice.  The main
problem consists in the size of the resulting set of equations: for
$n=3$ we have for each combination of $\pm$ for the 3 generating
reflections
% w.r.t. the planes, $x_2=1/2$, ..............
around 770 equation for the 256 coefficients $q_{D_1\ldots D_8}$;
for $n=4$ every set of equations for the coefficients of $Q_4$ has
around 250,000 equations for the 65536 coefficients  $q_{D_1\ldots
D_{16}}$. Naturally, not every combination of signs for the
generating reflections is possible, most of the resulting sets of
equations for  $q_{D_1\ldots D_{(2^n)}}$ allow only trivial solution
$q_{D_1\ldots D_{(2^n)}}=0$. The results of our computation are
given in table form in Theorem~\ref{th-fts} below.\footnote{To solve
the sparse but rather extensive linear systems for the coefficients
$q_{{\cal D}}$ appearing after the splitting w.r.t.\ the variables
$f_{i_1\ldots i_n}$, a special linear equation solver had to be
written. It can be downloaded together with other material related
to this publication (for details see {\tt
http://lie.math.brocku.ca/twolf/papers/TsWo2007/readme}).}

As it turns out, for $n=2$ three symmetry types of quasilinear
expressions $Q_2$ are possible. In the notation of Table
\ref{tablesymm} the first sign refers to the reflection on the line
$x_1=1/2$, the next sign stands for the reflection on the line
$x_1=x_2$. For example, the expressions of the first type $(+-)$ are
invariant  w.r.t.\ to the reflection on the line $x_1=1/2$, and show
a change of sign after the reflection on the line $x_1=x_2$. The
last case  $(++)$ consists of expressions which are invariant
w.r.t.\ any element of the complete group of symmetries of the
square which corresponds to the choice of the $(+)$ signs for the
two generating reflections of the square w.r.t.\ the lines $x_1=1/2$
and $x_1=x_2$.

For $n=3$ alongside with the completely symmetric quasilinear
expressions $Q_3$ (the third case $(+++)$ below), there are two
other cases $(---)$ and $(-++)$ of nontrivial quasilinear $Q_3$. In
this notation the first sign refers to the reflection on the plane
$x_1=1/2$, the next signs stand for reflections on the planes
$x_1=x_2$, $x_1=x_3$ (and $x_1=x_4$ for $n=4$).

The resulting numbers of free coefficients  $q_{{\cal D}}$
in the symmetric face formulas $Q_{n}$ are given
for each of the nontrivial cases in Table~\ref{tablesymm}.
We also give the number of nonzero terms in $Q_{n}$ for each case.

Especially remarkable is the totally skew-symmetric case $(---)$ for
$n=3$: it has only one (up to a constant multiple) nontrivial
expression
\begin{equation}\label{Q111}
\begin{array}{l}
   Q_{(---)}=(f_{100} - f_{001})(f_{010} - f_{111})(f_{101} -
f_{110})(f_{011} - f_{001}) - \\[0.5em]
  \qquad \qquad \qquad (f_{001} - f_{010})(f_{111} - f_{100})(f_{000} - f_{101})
   (f_{110} - f_{011}).
\end{array}
\end{equation}
Precisely this expression gives the so called discrete Schwarzian
bi-Kadomtsev-Petviashvili system (dBKP-system) --- an integrable
discrete system found in \cite{KSch,NSch} and studied in \cite{ABS}
where the fact of its $4d$-consistency was first established. The
dBKP-system has many equivalent forms and appears in very different
contexts. In addition to the known geometric interpretations and a
reformulation as Yang-Baiter system (\cite{Veselov}), the
dBKP-system may be considered as a nonlinear superposition principle
for the classical 2-dimensional Moutard transformations (\cite{GT}).

The expression (\ref{Q111}) enjoys an extra symmetry property: the
equation $Q_{(---)}=0$ is invariant under the action of the $SL_2({\mathbb C})$
group of fractional-linear transformations 
\begin{equation}
{\mathbf{f}}\longmapsto (a {\mathbf{f}} +b)
\big/(c {\mathbf{f}} +d)
\label{SL2}
\end{equation} 
(all group parameters $a$, $b$, $c$, $d$
are the same for all the vertices of the cube). This is in fact a direct
consequence of its uniqueness in this class. Since this  $SL_2({\mathbb C})$
action obviously preserves the symmetry type of an expression
w.r.t.\ the group action of the cube symmetry group, it is
reasonable to find the subclasses of  $SL_2$-invariant face
equations in each symmetry class. This is given in the third column
of the table. % Table~\ref{tablesymm}.
\begin{theorem}\label{th-fts}  The nonempty symmetry classes of
formulas $Q_n$ for face
dimensions $n=2,3,4$ are:
\begin{table}[ht]
 \begin{center}
\begin{tabular}{|c|l|l|} \hline
$n$ & types of symmetry, & number of parameters \\
    & number of parameters and terms & in $SL_2$-invariant subcases \\ \hline
 2  & 1) $({+}{-})$: 1 param.; \ 4 terms    & 1) 1 param.; 4 terms \\
    & 2) $({-}{+})$: 3 param.; 10 terms    & 2) none \\
    & 3) $({+}{+})$: 6 param.; 16 terms    & 3) 1 param.; 6 terms \\ \hline
 3  & 1) $({-}{-}{-})$: \ 1 param.; \ \ 24 terms & 1) 1 param.; \ 24 terms \\
    & 2) $({-}{+}{+})$: 13 param.; 186 terms & 2) none \\
    & 3) $({+}{+}{+})$: 22 param.; 256 terms & 3) 3 param.; 114 terms \\ \hline
 4  & 1) $({-}{-}{-}{-})$: \ 94 param.; $ 29208$ terms & 1) \ 5 param.;  15480 terms \\
    & 2) $({+}{-}{-}{-})$: \ 77 param.; $ 26112$ terms & 2) none \\
    & 3) $({-}{+}{+}{+})$: 349 param.; $ 60666$ terms & 3) \ 3 param.;  15809 terms \\
    & 4) $({+}{+}{+}{+})$: 402 param.; $2^{16}$ terms & 4) 18 param.;  96314 terms \\ \hline
\end{tabular}
 \caption{Symmetry classification of the face formulas w.r.t\ the
 complete symmetry group of the cube}
 \end{center}
\end{table}  \label{tablesymm}
\end{theorem}

In particular, the explicit expression for the
(non-$SL_2$-symmetric) $2d$-face formulas are:
\begin{eqnarray}
 (+-):& Q=&q_1 (f_{11} f_{10} - f_{11} f_{01}
   - f_{10} f_{00} + f_{01} f_{00})
    = q_1 (f_{11} - f_{00})(f_{10} - f_{01}), \label{2dpm}\\
 (-+):& Q=&
%\begin{array}
q_1 (f_{11} - f_{10} - f_{01} + f_{00})
  + q_2 (f_{11} f_{00} - f_{10} f_{01})
  +  \label{2dmp}\\
 & &q_3 (f_{11} f_{10} f_{01} - f_{11} f_{10} f_{00} - f_{11}
f_{01} f_{00}
  + f_{10} f_{01} f_{00}),\nonumber \\
%\end{array}\\
%\begin{eqnarray*}
 (++): & Q=&  q_1 +
 q_2 (f_{11} + f_{10} + f_{01} + f_{00})
        + q_3 (f_{11} f_{00} + f_{10} f_{01}) + \label{2dpp} \\
& &     q_4 (f_{11} f_{10} + f_{11} f_{01} + f_{10} f_{00}
+ f_{01} f_{00}) +  \nonumber \\
 & &q_5 (f_{11} f_{10} f_{01} + f_{11} f_{10} f_{00}
+ f_{11} f_{01} f_{00} + f_{10} f_{01} f_{00}) + \nonumber  \\
 & &     q_6 f_{11} f_{10}f_{01} f_{00}. \nonumber
\end{eqnarray}
The explicit form of the $SL_2$-symmetric face formula for $n=2$ in
the first $(+-)$ case is the same:
$$Q=q_1 (f_{11} - f_{00})(f_{10} - f_{01}).
$$
For $n=2$ and the symmetric case $(++)$ the $SL_2$-symmetric face
formula is:
$$Q=q_1(f_{11} f_{10} + f_{11} f_{01} - 2 f_{11} f_{00} - 2 f_{10} f_{01}
   + f_{10} f_{00} + f_{01} f_{00}).
$$
It should be noted that (\ref{2dmp}) is $3d$-consistent; it can be
simplified using $SL_2$ transformations and put into explicitly
linear form $Q=f_{11} - f_{10} - f_{01} + f_{00}$ or into $Q=f_{11}
f_{00} - f_{10} f_{01}$, so the respective equation $Q=0$ is
equivalent to a linear equation $\log f_{11} + \log f_{00} = \log
f_{10} + \log f_{01}$.

\section{Nonexistence of nontrivial integrable face formulas in other
         symmetry classes for $n=3$}\label{sec4}
In this section we give a computational proof of our main result:
\begin{theorem}\label{th-nonex}  Among the three possible symmetry types
of $3$-di\-men\-si\-o\-nal quasilinear face formulas given in the second
column of Table~\ref{tablesymm}, only formula (\ref{Q111})
gives a non-trivial $4d$-compatible face equation. Any $4d$-compatible
face formula in the other two symmetry types may be transformed using the
action of the group  $SL_2({\mathbb C})$ on the field variables to one of the following
linearizable forms:
\begin{equation}\label{cases1}
Q^{(1)}=f_{000}f_{001}f_{010}f_{011}f_{100}f_{101}f_{110}f_{111} -\sigma,
\end{equation} 
\begin{equation}\label{cases2}
Q^{(2)}=f_{001}f_{010}f_{100}f_{111} -\sigma f_{000}f_{011}f_{101}f_{110}, \end{equation} 
\begin{equation}\label{cases3}
Q^{(3)}=(f_{001} + f_{010} + f_{100} + f_{111}) -\sigma
          (f_{000} + f_{011} + f_{101} + f_{110}), 
\end{equation} 
where $\sigma=\pm 1$. 
\end{theorem}
Technically,  in order to find $4d$-consistent $3d$-face formulas among
the other two %nontrivial
$3$-dimensional cases $(-{+}{+})$ and $({+}{+}{+})$ (as listed in Table~\ref{tablesymm}),
one shall run the following algorithmic steps:

\textit{Step 1}. Take a copy of this $Q_3$ formula, map it onto the four
initial faces of the $4d$-cube (where one of the coordinates $x_i=0$), solve the
mapped equations
with respect to $f_{0111}$,  $f_{1011}$, $f_{1101}$ and $f_{1110}$
leaving the other variables free.

\textit{Step 2}.  Then substitute the obtained \textit{rational}
expressions for $f_{0111}$,  $f_{1011}$, $f_{1101}$ and $f_{1110}$
into the copies of the face formula mapped onto the four
final faces (where  one of the coordinates $x_i=1$), finding
respectively four different expressions for the last vertex field
$f_{1111}$.

\textit{Step 3}.  Equate these 4 expressions for  $f_{1111}$
obtaining three rational equations in terms of 11 free variables $f_{0000}$,
$f_{0010}$, $f_{0101}$, \ldots \  and
the parametric coefficients  $q_{D_1\ldots D_8}$ left free in the
chosen symmetry class.

\textit{Step 4}. Removing the common denominators of the equations
and splitting the resulting polynomials w.r.t.\ the 11 free variables
$f_{ijkl}$ one obtains a polynomial system of equations for the free
coefficients   $q_{D_1\ldots D_8}$.

\textit{Step 5}. The latter should be solved, resulting in a complete
classification of $4d$-consistent quasilinear scalar $3d$-face formulas.

This approach, applied in a straightforward way, results in
extremely huge expressions. Even building the rational expressions
in Step~4 in a straightforward way seems to be unrealistic: for a
typical $3d$-face formula from Table~\ref{tablesymm} Step~4 should
end up (as our test runs allowed for an estimate) in an expression
with around $10^{14}$ terms, which is beyond the reach of computer
algebra systems in the foreseeable future. Even brute force testing
of $4d$-consistency of the smallest solution (face formula
(\ref{Q111}) which has no free parametric coefficients $q_{D_1\ldots
D_8}$) results in $\approx 2\cdot 10^{8}$ terms (after substituting
the expressions for $f_{0111}$, $f_{1011}$, $f_{1101}$, $f_{1110}$,
collecting the terms over the common denominator and expanding the
brackets before the cancellation can start in Step~4\footnote{The
commented {\sc Form} logfile for this estimate can be
downloaded from \\
{\tt
http://lie.math.brocku.ca/twolf/papers/TsWo2007/BruteForceCheck/Case---/}}).
Technically this is explained by the presence of 4 different
symbolic denominators of the rational expressions for $f_{0111}$,
$f_{1011}$, $f_{1101}$, $f_{1110}$ and their various products. A
careful step-by-step substitution and cancellation of like terms in
several stages still can be done even on a modest computer \emph{for
this $(---)$ case}. Using the system {\sc Form} (this system was
specially designed for large symbolic computations), one can prove
that all terms finally cancel out for the case of the integrable
$3d$-face formula (\ref{Q111}) thus giving a computational proof of
its $4d$-consistency in 3 min CPU time (3~GHz Intel running Linux
SUSE 9.3) and less than 200 Mb disk space for temporary data
storage.\footnote{The {\sc Form} code of this run and its logfile
can be
downloaded from \\
{\tt http://lie.math.brocku.ca/twolf/papers/TsWo2007/BruteForceCheck/Case---/}).}

As the estimates given in Appendix~A show, the straightforward
approach based on Steps~1--4 is unrealistic for the other two
3-dimensional cases listed in Table~\ref{tablesymm}, even at the stage
of generation of the consistency conditions (Step~4).

In order to classify discrete integrable $3d$-face formulas $Q_3=0$
for the case $(-{+}{+})$ and the hardest case $({+}{+}{+})$ we used a
totally different randomized ``probing'' strategy, explained in detail
in Appendices~B,~C,~D.

After the computation (cf.\ Appendices~B--E) the list of candidate
face formulas $Q_3$ for the case $({+}{+}{+})$ included 5 formulas
(before the verification that these formulas, obtained by our
``probing'' method, really give $4d$-consistent $3d$-formulas).  For
the case $(-{+}{+})$ the list of candidate face formulas included 3
formulas. All of them include a few free parameters.
As one can show,
all these formulas can be greatly simplified using the action of the group
$SL_2({\mathbb C})$ on the field variables ${\mathbf f}$, resulting in the $4d$-consistent $3d$-face formulas (\ref{cases1}), (\ref{cases2}),
(\ref{cases3}).
The first two formulas  (\ref{cases1}), (\ref{cases2})
can be linearized using the logarithmic substitution
$\tilde{f}_{ijk}= \log f_{ijk}$.

The expressions for the aforementioned candidates, the {\sc Form}
procedures and their logfiles showing the simplification process can
be downloaded from \\
{\tt http://lie.math.brocku.ca/twolf/papers/TsWo2007/}. Here we
just give one example of such a $SL_2$-simplification: the $Q$-expression for the case $Q1$ in \\
{\tt http://lie.math.brocku.ca/twolf/papers/TsWo2007/SL2-simplification/Case+++/} is
\begin{eqnarray*}
Q&=&q_{105}\big(f_{001}f_{010}f_{100}f_{111} + f_{000}f_{011}f_{101}f_{110}\big) + \\
 & &q_{107}\big(f_{001}f_{010}f_{100}f_{110}f_{111} + f_{001}f_{010}f_{100}f_{101}f_{111} + f_{001}f_{010}f_{011}f_{100}f_{111}+ \\
 & &\ \ \ \ \ \ f_{000}f_{011}f_{101}f_{110}f_{111} + f_{000}f_{011}f_{100}f_{101}f_{110} + f_{000}f_{010}f_{011}f_{101}f_{110} + \\
 & &\ \ \ \ \ \ f_{000}f_{001}f_{011}f_{101}f_{110} + f_{000}f_{001}f_{010}f_{100}f_{111}\big) + \\
 & &\frac{q_{107}^2}{q_{105}}\big(f_{001}f_{010}f_{100}f_{101}f_{110}f_{111} + f_{001}f_{010}f_{011}f_{100}f_{110}f_{111} + \\
 & &\ \ \ \ \ \                   f_{001}f_{010}f_{011}f_{100}f_{101}f_{111} + f_{000}f_{011}f_{100}f_{101}f_{110}f_{111} + \\
 & &\ \ \ \ \ \                   f_{000}f_{010}f_{011}f_{101}f_{110}f_{111} + f_{000}f_{010}f_{011}f_{100}f_{101}f_{110} + \\
 & &\ \ \ \ \ \                   f_{000}f_{001}f_{011}f_{101}f_{110}f_{111} + f_{000}f_{001}f_{011}f_{100}f_{101}f_{110} + \\
 & &\ \ \ \ \ \                   f_{000}f_{001}f_{010}f_{100}f_{110}f_{111} + f_{000}f_{001}f_{010}f_{100}f_{101}f_{111} + \\
 & &\ \ \ \ \ \                   f_{000}f_{001}f_{010}f_{011}f_{101}f_{110} + f_{000}f_{001}f_{010}f_{011}f_{100}f_{111}\big) + \\
%\end{eqnarray*}
%\begin{eqnarray*}
 & &\frac{q_{107}^3}{q_{105}^2}\big(f_{001}f_{010}f_{011}f_{100}f_{101}f_{110}f_{111} + f_{000}f_{010}f_{011}f_{100}f_{101}f_{110}f_{111} + \\
 & &\ \ \ \ \ \                     f_{000}f_{001}f_{011}f_{100}f_{101}f_{110}f_{111} + f_{000}f_{001}f_{010}f_{100}f_{101}f_{110}f_{111} + \\
 & &\ \ \ \ \ \                     f_{000}f_{001}f_{010}f_{011}f_{101}f_{110}f_{111} + f_{000}f_{001} f_{010}f_{011}f_{100}f_{110}f_{111} + \\
 & &\ \ \ \ \ \                     f_{000}f_{001}f_{010}f_{011}f_{100}f_{101}f_{111} + f_{000}f_{001}f_{010}f_{011}f_{100}f_{101}f_{110}\big) + \\
 & &2\frac{q_{107}^4}{q_{105}^3}\big(f_{000}f_{001}f_{010}f_{011}f_{100}f_{101}f_{110}f_{111}\big)
\end{eqnarray*}
The simplifying transformation consists  in the following steps:
\begin{enumerate} % \itemsep=-3pt
% 1
\item $f_{ijk} \mapsto  \frac{q_{105}}{q_{107}} f_{ijk}$ and  $Q\mapsto \frac{q_{107}^4}{q_{105}^5} Q$
(this eliminates the parametric $ q_{105}$ and $ q_{107}$)
% 1

\item $f_{ijk} \mapsto  \frac{1}{f_{ijk}}$ (and removing the denominator in $Q$ afterwards)

% 1
\item $f_{ijk} \mapsto f_{ijk}-1$.
\end{enumerate}
This produces the simplified form $Q=f_{001}f_{010}f_{100}f_{111}  + f_{000}f_{011}f_{101}f_{110}$.

\section*{Acknowledgments}

For this work facilities of the Shared Hierarchical Academic Research
Computing Network (SHARCNET: {\tt www.sharcnet.ca}) were used.

TW thanks the Konrad Zuse Institut at  Freie Universit\"{a}t Berlin
and the Techni{sch}e Universit\"{a}t Berlin where part of the work was done.

\appendix

\section*{Appendix A: The Size of Consistency Conditions}\label{consistency}  %=============

The following considerations are made under the assumption
of generic unknown coefficients $q_{{\cal D}}$ in the face formula
(\ref{gf}) %#######
not satisfying additional symmetry conditions.

Any $(n+1)$-dimensional hypercube built from $2^{n+1}$ vertices
$f_{i_1\ldots i_{n+1}}$, $ i_k\in\{0,1\}$ has $2(n+1)$ faces located in the
(logical) planes $x_k=0$ and $x_k=1$, $ k=1,\ldots, (n+1)$.
The face relations for the $n+1$ faces that correspond to $x_k=0$ are
\begin{equation}\label{0faces}
0 % = Q_n|_{0 \rightarrow k}
=\sum_{{\cal D}} q_{{\cal D}} \prod_{i_s = 0,1}
    \big(f_{i_1\ldots i_{k-1} 0\,i_{k+1}\ldots i_{n+1}}\big)^
    {D_{i_1\ldots i_{k-1}i_{k+1}\ldots i_{(n+1)}}}.
\end{equation}
They can be used to
determine $f_{1..101..1}$ with the 0 being in the
$k^{\rm th}$ index position.
Each of these face relations involves $2^n$ $f$-variables
% $f_{i_1\ldots i_{k-1} 0 i_{k+1}\ldots i_{n+1}}$
and thus $2^{2^n}$ terms, half of them include $f_{1..101..1}$ as a factor
and the other half not. Solving the face
relation $x_k=0$:
\begin{equation}
0 = A_k f_{1..101..1}+B_k   \label{zk0fr}
\end{equation}
where $A_k$, $B_k$ are expressions in $q_{{\cal D}}, f_\beta$ for
$f_{1..101..1}$ and substituting $f_{1..101..1}=-B_k/A_k$ in any
expression that involves $f_{1..101..1}$ linearly (like other face
relations) and taking the numerator over the common denominator
amounts to multiplying all terms that involve $f_{1..101..1}$ by
$-B_k$ and all other terms by $A_k$. As $A_k$ and $B_k$ involve each
$2^{2^n}/2=2^{2^n-1}$ terms this means that a substitution of
$f_{1..101..1}$ increases the number of terms by a factor of
$2^{2^n-1}$, before cancellations and reductions will be made.

The $2^{\rm nd}$ half of face relations for the  $n+1$ faces that
correspond to $x_k=1$ are
\begin{equation}\label{1faces}
0 % = Q_n|_{1 \rightarrow k}
=\sum_{{\cal D}} q_{{\cal D}} \prod_{i_s = 0,1}
    \big(f_{i_1\ldots i_{k-1} 1\,i_{k+1}\ldots i_{n+1}}\big)^
    {D_{i_1\ldots i_{k-1}i_{k+1}\ldots i_{(n+1)}}}.
\end{equation}
Each one of them involves $f_{11..1}$ and $n$ of those % $n+1$
$f$-variables which have exactly one 0 as index in any one of the
$n+1$ index positions apart from the $k^{\rm th}$ position. Replacing
each one of these $n$\ $f$-variables by using the corresponding
$x_l=0$ %$Q_n|_{0\rightarrow j}$
face relation increases the number of terms
% of $Q_n|_{1 \rightarrow k}$
by a factor $2^{2^n-1}$ each time, giving in
total $2^{2^n}\big(2^{2^n-1}\big)^n = 2^{2^n(n+1)-n}$ terms.  In each
substitution the degree of the coefficients $q_{{\cal D}}$ increases
by one, reaching finally $n+1$.

Solving one of the $n+1$ many $x_k=1$ face relations
\begin{equation}
0 = G_k f_{11..1}+H_k   \label{zk1fr}
\end{equation}
for $f_{11..1}$ and substituting $f_{11..1}=-H_k/G_k$
in the other face relations gives $n$ independent
consistency conditions
\begin{equation}
G_jH_k=G_kH_j, \ \ \ j=1,\ldots,k-1,k+1,\ldots,n+1  \label{concon}
\end{equation}
with each $G_i$ and $H_i$
having $2^{2^n(n+1)-n}/2$ terms, i.e.\ each consistency condition
involving $2\big(2^{2^n(n+1)-n-1}\big)^2$ terms. The total number of
terms of the $n$ consistency conditions is thus $n 2^{\{2^{n+1}(n+1)-2n-1\}}$.

To compute an upper bound of the number of conditions that result from
splitting each consistency condition with respect to the independent
$f$-variables we note that their highest degree is equal to the total
degree of all $q_{{\cal D}}$, i.e.\ it is $2n+2$. The only exception
is $f_{00\ldots 0}$ which does not occur in the $0 = Q_n|_{1
\rightarrow k}$ face relations and enters only through substitutions,
so its highest degree in the constraints is $2n$. We thus get for an
upper bound of the number of different products of different powers of
$f_{00\ldots 0}$ and the other
$2^{n+1}-n-3$ independent $f$-variables the value
$2n(2n+2)^{\big(2^{n+1}-n-3\big)}$. With this number and the number of
terms of each constraint we get with their quotient %a lower bound on
an estimate of the
average number of terms in each equation (see Table~2). %\ref{Ttable}).

\begin{table}
 \begin{center}
\scriptsize
 \begin{tabular}{|l|c|c|c|c|c|} \hline
$\begin{array}{l}
\mbox{dimension of face}
\end{array} $      & $n$           & 2 &  3 &  4 &  5  \\ \hline
$\begin{array}{l}
\mbox{\# of $f$-variables in face formula}
\end{array} $      & $2^n$         & 4 &  8 & 16 & 32  \\ \hline
$\begin{array}{l}
\mbox{\# of terms in face formula} \\
\mbox{(= \# of undetermined} \\
\mbox{coefficients
$q_{{\cal D}}$ in $Q_n$ })
\end{array} $      & $2^{2^n}$     &16 &256 &65536 & $4.3\times10^9$ \\ \hline
$\begin{array}{l}
\mbox{\# of all $f$-variables} \\
\mbox{in $(n+1)$-dim.\ hypercube}
\end{array} $      & $2^{n+1}$     & 8 & 16 & 32 & 64  \\ \hline
$\begin{array}{l}
\mbox{\# of indep.\ $f$-variables} \\
\mbox{in $(n+1)$-dim.\ hypercube}
\end{array} $      & $2^{n+1}-n-2$ & 4 & 11 & 26 & 57  \\ \hline
$\begin{array}{l}
\mbox{\# of $n$-dim.\ faces} \\
\mbox{in $(n+1)$-dim.\ hypercube}
\end{array} $      & $2(n+1)$      & 6 &  8 &  10 &  12  \\ \hline
$\begin{array}{l}
\mbox{\# of consistency conditions}
\end{array} $      & $n$           & 2 &  3 &  4 &  5  \\ \hline
$\begin{array}{l}
\mbox{upper bound on the \# of terms}\\
\mbox{of each condition}
\end{array} $      & $2^{\{2^{n+1}(n+1)-2n-1\}}$
                   & $5.2\times 10^5$    & $1.4\times 10^{17}$
                   & $2.8\times 10^{45}$ & $1.9\times 10^{112}$ \\ \hline
$\begin{array}{l}
\mbox{total degree of the $q_{{\cal D}}$} \\
\mbox{in each condition}
\end{array} $      & $2n+2$        & 6 &  8 & 10 & 12  \\ \hline
$\begin{array}{l}
\mbox{upper bound estimate of the}        \\
\mbox{\# of equations resulting from}     \\
\mbox{splitting each condition }
\end{array} $      & $2n(2n+2)^{\big(2^{n+1}-n-3\big)}$
                   & 864  & $6.4\times10^9$ & $8.0\times10^{25}$
                   & $2.7\times10^{61}$  \\ \hline
$\begin{array}{l}
\mbox{estimated average \# of terms}      \\
\mbox{in each equation}
\end{array} $      & $\frac{2^{2^{n+1}(n+1)-2n-1}}
                           {2n(2n+2)^{\big(2^{n+1}-n-3\big)}}$
                   & 606 & $2.2\times 10^7$
                   & $3.5\times 10^{19}$ & $7\times 10^{50}$  \\ \hline
 \end{tabular}
 \caption{Some statistics of faces and consistency conditions}
 \end{center}
\end{table}  \label{Ttable}
\normalsize

\textit{Remark.}  Although, strictly speaking, these are upper
bounds for the size of conditions, one shall keep in mind that any
computer algebra system \textit{shall generate all these terms}
expanding the brackets and only after generating all of them or a
part of them it can search for possible cancellations or reductions.
As our test runs with {\sc Form} had shown, for the 3-dimensional
case $(+++)$ given in Table~\ref{tablesymm} (with a much smaller
number of independent $q_{{\cal D}}$ than $2^{2^n}$ but the same
number of terms in $Q_3$), the total number of terms to be generated
for each of the 3 consistency conditions
is around $10^{14}$ (compared to $1.4\cdot 10^{17}$ in Table~2).  %\ref{Ttable}).
A typical single equation for the coefficients  $q_{{\cal D}}$ % which are free in this case $(++)$
resulting from splitting a partially formed consistency condition
has a few thousand terms of degree 8 in the parametric $q_{{\cal D}}$.

\section*{Appendix B: The computational problem} \label{TheProblem}  %=============
As outlined before (cf.\ Section~\ref{sec4}), the task consists of the following steps.
\begin{enumerate} \itemsep=-3pt
% 1
\item Formulate the relations for the $n+1$ faces $x_k=0$.
% 2
\item Solve them for $f_{11\ldots 101\ldots 1}$.
% 3
\item Formulate the relations for the $n+1$ faces $x_k=1$.
% 4
\item Perform the substitutions obtained under 2.\ in the
      relations of 3.
% 5
\item Solve one of the resulting relations for $f_{11\ldots 1}$ and
% 6
\item Substitute $f_{11\ldots 1}$ in all other $n$ face relations of 4.
% 7
\item Split these consistency conditions with respect to all the occurring
      independent $f$-variables to obtain an overdetermined
      system of equations for the unknowns $q_{{\cal D}}$.
% 8
\item Find the general solution of this system.

\item Reduce the number of free parameters of the solutions using
       $SL_2({\mathbb C})$-transformations (\ref{SL2}).

\end{enumerate}

Although the algebraic system for the unknown coefficients $q_{{\cal D}}$
is heavily overdetermined the following difficulties appear.
\begin{enumerate}\itemsep=-3pt
% 1
\item Strictly speaking, in order to formulate even only the smallest
subset of conditions one would have
to formulate at least one consistency condition (by performing steps
1., 2.\ fully and 3.- 6.\ for at least two $x_k=1$
face relations (\ref{1faces}) before step 7)  i.e.\ to generate an expression with
$2^{\{2^{n+1}(n+1)-2n-1\}}$ terms.
% 2
\item If one found a way around this hurdle then the resulting equations
are of high degree $2n+2$ with on average many terms.
% 3
\item Even if one were able to generate 100,000's of equations and
thus find shorter ones which one could solve for some unknowns in
terms of others, one would face the problem that many cases and
sub-sub-cases have to be investigated due to the high degree of the
equations and
% 4
\item that one has to generate  billions of equations to find some
that are independent of the ones generated so far.
As we explain below, one can not hope that the first, say, $10^6$ conditions
will be equivalent to the full system of equations for $q_{{\cal D}}$, even though
we have only very few unknown $q_{{\cal D}}$'s due to the ``triangular'' form of this huge
system as we explain in Appendix~D. %Appendix~\ref{probing}.
\end{enumerate}

In the computations to be described in this paper we have $n=3$.  The
full problem (without cubical symmetry) would mean to generate 3 consistency conditions involving
each about $10^{17}$ terms that split into an estimate of $10^{10}$
polynomial equations each being homogeneous of degree 8 for 256
unknowns and involving on average over $10^7$ terms.

To make progress we introduce different solving techniques but also
simplify the problem:
\begin{enumerate}
\item
We restrict our problem to face formulas that obey the full cubical
symmetry  (cf.\ Section~\ref{sec3}). 
This reduces the problem to 3 cases: 

$\bullet$ case $(---)$ with 232
identically vanishing $q_{{\cal D}}$ and 24 other  $q_{{\cal D}}$  depending
on 1 parameter, 

$\bullet$ case $(-++)$ with 70 identically vanishing
$q_{{\cal D}}$ and 186 other $q_{{\cal D}}$ depending on 13 parameters,
and 

$\bullet$ case $(+++)$ with no identically vanishing $q_{{\cal D}}$ and
all 256 $q_{{\cal D}}$ depending on 22 parameters. 

In the third case
$(+++)$ -- the hardest case -- the fact that only 22 of the
$q_{{\cal D}}$ are parametric simplifies the problem but the
simplification is limited because none of the other 234  
$q_{{\cal D}}$ needs to
vanish just because of the extra cubical symmetry (see
Section~\ref{sec3}).
\item
We ease the formulation of the large consistency conditions (before
splitting them into smaller equations) by replacing $z$ of the
$v:=2^{n+1}-n-2$ independent $f$-variables by {\em z}ero, and replacing $u$
of them by random integer values {\em u}n-equal zero, leaving $s:=v-z-n$ of them in
{\em s}ymbolic form.  As a consequence, we only get a set of necessary and
not sufficient equations for the $q_{{\cal D}}$ but we can repeat this
procedure on a gradually increasing level of generality (by increasing $s$
and lowering $z$). A crucial
feature is that the preliminary knowledge about the solution is used
in the formulation of new necessary systems of equations
(see Appendix~C).  %\ref{probing}).
\item
We design a dynamic process that automatically organizes an iteration
process that generates and solves/simplifies/'makes use of' necessary
equations automatically. As an important feature, a set of newly
generated necessary equations is not read into the ongoing computation
at once but gradually on demand (see Appendix~D). %\ref{cache}).
\item
A recent extension is the parallelization of different case
investigations on a computer cluster. % (see section \ref{parallel}).
\end{enumerate}

We discuss points 1-3 now in more detail.

\section*{Appendix C: Random probing%Probabilistic conditions
}  \label{probing}  %=============
In this Appendix we explain the ``probing technique'' in more detail
which was mentioned briefly in Appendix~B
under point 2.\ in the list of techniques.
It is based on replacing a number $z$ of the
$v:=2^{n+1}-n-2$ independent $f$-variables by zero
(for $n=3$,  $v=11$), and replacing a number $u$
of them by random integer non-zero values,
leaving $s:=v-z-u$ of them in symbolic form.

The computation has 3 phases: finding solutions,
verifying the obtained solutions probabilistically
and again rigorously.
The first two phases use the probing technique.
Still, it makes sense to distinguish them because
optimal values of 
$z, u$ and $s$
%the optimal numbers of independent $f_\alpha$ (hereafter we use the
%shortcut $\alpha=(i_1 \ldots i_{n+1})$, $i_s=0,1$
%for the indices of the field variables)
%to be replaced by zero and by non-zero integers
differ in both cases.

In the probing technique the two types of replacements, 
i.e.\ replacing $f_\alpha$ by zero or by a non-zero integer, share the
same disadvantage: the number of independent parameters, i.e.\ of
symbolic $f_\alpha$ which allow to split the consistency condition
into many smaller equations, is reduced by one.

The advantage of replacing an $f_\alpha$ by zero instead of a non-zero
integer is that expressions shrink more. Also, replacements by
non-zero integers often introduce extra solutions to the generated
conditions and it appears to be costly to eliminate these spurious
temporary solutions by leading them to a contradiction with more
conditions to be generated based on other random
replacements. Therefore it is more productive to replace, for example,
{\em one} $f_\alpha$ by zero than to replace {\em two} $f_\alpha$ by
non-zero integers.

Consequently, in the first phase of {\em finding solutions} at most
one replacement by a non-zero integer is made (i.e.\ $u \leq 1$) and
after starting with $z=9, u=0, s=2$ one increases generality gradually
by either changing $u$ from 0 to 1 and decreasing $z$ by 1, or by
decreasing $u$ from 1 to 0 and increasing $s$ by 1 until $z=1, u=0,
s=10$.  We need the occasional substitution by one non-zero integer
because we want to increase the generality in as small as possible
steps in order to avoid the generation of too many too high degree
equations with too many terms. This would happen if we decrease $z$ by
1, keep $u=0$ and increase $s$ by 1.  A run in full generality $z=0,
u=0, s=11$ is computationally prohibitive, therefore in a second
phase of {\em confirming the found solution probabilistically} one
starts with $z=4, u=1, s=6$ and increases generality by decreasing
$z$, increasing $u$ and keeping $s$ constant until $z=0, u=5, s=6$.
By testing a hypothetical solution with this final setting
many times, the correctness of the solution is confirmed with an
arbitrarily high probability.

In both phases we want to generate as few and as simple equations as
possible in each generation step. So we continue using the same
setting of $z,u,s$ as long as possible (i.e.\ as long as there are still
resulting new conditions after randomly choosing other sets of $z$ many
$f_\alpha$ to be 0, $s$ of $f_\alpha$ to be kept symbolic and randomly
assigning integer non-zero values to the other $u$ parametric $f_\alpha$)
before generalizing it, i.e.\ making $s$ larger and $z$ smaller.
This is regulated by one parameter which specifies the maximum number of
consecutive times that a `probing' (generation of conditions) attempt
yielded only identities before changing $z,u,s$.

In a third phase, after all hypothetical solutions have been obtained
and been checked probabilistically, they are checked again, now {\em
rigorously}. This has been done either by a brute force check using
the computer algebra system {\sc Form} or by using
$SL_2({\mathbb C})$-transformations on the field variables ${\mathbf f}$
to reduce solutions to integrable trivial forms (\ref{cases1}),
(\ref{cases2}), (\ref{cases3}) in Section~\ref{sec4}.

A helpful and initially unexpected feature of the probing technique is
that the resulting equations appear to be somehow triangularized in
the following sense. Each unknown
$q_{i_1i_2\ldots i_m}$,  $m=2^{n}$,  $i_j \in \{0,1\}$
is the coefficient of a product of $i_1+i_2+\ldots +i_m$ many
different factors $f_\alpha$. That means, that at the beginning of the
computation when many $f_\alpha$ are replaced by zero, the $q_{{\cal
D}}$ with a high index sum do not occur. Only later on as fewer
$f_\alpha$ are replaced by zero, gradually $q_{{\cal D}}$ with higher
index sum appear in the equations. On one hand this is a good feature,
providing a partially triangularized system of equations. On the other
hand this means that although we only want to compute a relatively
small number of unknowns (for $n=3$ and case (+++) these are 22 $q_{{\cal
D}}$) it is not enough to formulate only a comparable number of the
huge total set of equations (about $6.4\times10^9$ equations for $n=3$).
A set of equations that is equivalent to the complete set of equations is
only obtained towards the end of generalizations. For example,
$q_{11\ldots 1}$ as the coefficient of the product of all the $2^n$
many $\mathbf{f}$ occurring in at least one face formula, at most $2^n-1$ of
them belonging to the independent $f_\alpha$, can only occur in at least one
consistency condition
if none of the $f_\alpha$ occurring in at least one face formula is
replaced by zero, i.e.\ if $u$ and $s$ are big enough to satisfy $(u+s)
\geq (2^n-1)$.

That in turn means that millions of the early equations are redundant
which implies a large inefficiency in generating equations. This can
be avoided by using known relations
$q_{{\cal D}}=h_{{\cal D}}(q'_{{\cal D}})$,
which were derived in the solution process so far, as automatic
simplification rules when generating new equations.

Three problems remain to be considered.

1. By replacing $f_\alpha$ through numbers it may happen that $A_k$ in
any one of the $x_k=0$ face relations (\ref{zk0fr}) becomes zero. Then a
new equation generation attempt with different or more general random
replacements has to be made.

2. Similarly, it may happen that the coefficient of $q_{11\ldots 1}$
in all $n+1$ many $x_k=1$ face relations (\ref{zk1fr}) is zero. Then a
different replacement has to be tried as well.

3. Even if none of the $A_k$ in (\ref{zk1fr}) becomes zero it may
happen that $A_k$ and $B_k$ in one face relation are not prime and
then a solution for the $q_{{\cal D}}$ which makes the greatest common
divisor $GCD(A_k,B_k)$ to zero is potentially lost when performing
substitutions $f_{1..101..1}=-B_k/A_k$. The same applies to common
factors of $G_k$ and $H_k$ in the single face relation (\ref{zk1fr})
that is applied to replace $f_{11\ldots 1}$. Therefore each
consistency condition
% resulting from a substitution of $f_{11\ldots 1}$
has to be multiplied with a product of all common factors of any
pair $A_k$, $B_k$ and of all common factors of the pair $G_k$, $H_k$ which
is used to replace $f_{11\ldots 1}$. To lower the computational cost
one drops multiplicities of the factors. These factors involve in general
$q_{{\cal D}}$ as well as $f_\alpha$ and therefore the multiplication has to
be done before splitting the consistency condition with respect to the
independent $f_\alpha$. Alternatively one can split the consistency
condition before multiplication and instead multiply and duplicate the
equations in the following way.

Let $P(q_{{\cal D}},f_\alpha)$ be one of the above mentioned factors and
let $0=P$ be split into a system $0=\hat{P}_i(q_{{\cal D}})$ where redundant
equations are dropped.\footnote{The definition of `redundant' depends
of the effort one wants to spend at this stage. In the implementation
of this algorithm the polynomials $\hat{P}_i$ are divided by the coefficients
of their leading terms with respect to some ordering of the $q_{{\cal D}}$
and then duplicate $\hat{P}_i$ are dropped.} Instead of multiplying a constraint
$0=C(q_{{\cal D}},f_\alpha)$
with $P$, splitting the equation $0=PC$ into individual equations
and factorizing all of them afterwards, it is equivalent but much more efficient
to split $0=C$ into a system of equations $0=\hat{C}_j(q_{{\cal D}})$
and $0=P$ into a system $0=\hat{P}_i(q_{{\cal D}})$ and to consider the
equivalent system $0=\hat{P}_i\hat{C}_j$, $ \forall i,j$.

To summarize, in order not to loose solutions for the $q_{{\cal D}}$, the
procedure is
\begin{itemize}
\item to collect all common factors $P_r$ of all pairs $A_k$, $B_k$ and
of the pair $G_k$, $H_k$ used to substitute $f_{11\ldots 1}$,
\item to drop duplicate factors,
\item to split all consistency conditions giving a system $S$ of
equations $0=\hat{C}_j$, and
\item to split for each factor $P_r$ the equation $0=P_r$ into a
system $0=\hat{P}_{ri},\ \ i=1\ldots i_r$ where again
equations that are redundant within one such system are dropped.
\item If a system $0=\hat{P}_{ri}$ includes a non-vanishing
$\hat{P}_{ri}$ either because $\hat{P}_{ri}=1$ or because
$\hat{P}_{ri}(q_{{\cal D}})$ is
known to be non-zero based on the inequalities that are known for some
$q_{{\cal D}}$ then this system is ignored because the corresponding $P_r$
is non-zero.  For every other such system $0=\hat{P}_{ri},$
replace the system $S$ of conditions $0=\hat{C}_j$ by the new system
$\hat{S}$ consisting of the equations $0=\hat{P}_{ri}\hat{C}_j$, $ \forall i,j$.
\end{itemize}

\section*{Appendix D: A Succession of Generating and
          Solving Equations} \label{cache} %=============
The system of algebraic equations for the $q_{{\cal D}}$ is
investigated by the computer algebra package {\sc Crack} that aims at
solving polynomially algebraic or differential systems, typically
systems that are overdetermined and very large. It offers various
degrees of interactivity from fully automatic to fully interactive.
The package consists of about 40 modules which perform different
steps, like substitutions, factorizations, shortenings, Gr\"{o}bner
basis steps, integrations, separations,... which can be executed in
any order. In automatic computations their application is governed by
a priority list where highly beneficial, low cost and low risk (of exploding the size of
equations) steps come first.  Modules are tried from the beginning of
this list to its end until an attempt is successful and then execution
returns to the start of the list and modules are tried again in that
order. This simple principle is refined in a number of ways. For more
details see \cite{CrackOver}, \cite{CrackOnline}.

In order to accommodate the dynamic generation of equations and
their successive use, all that had to be done was  to add two modules:
\begin{itemize}
\item one for generating a new set of necessary
conditions using the `probing' technique from Appendix~C %\ref{probing}
and writing the generated equations into a buffer file, and
\item another module for reading one non-trivial equation from this buffer
file (i.e.\ for continuously reading equations until one is obtained
that is not instantly simplified to an identity modulo the known
equations or until the end of this file is reached), 
\end{itemize}
and to determine the place of these modules in the priority list.
The need for a buffer file arose because the number of equations
generated in each `probing' is unpredictable, especially in view of
the large impact that factors $\hat{P}_{ri}$ can have on the number of
equations (see the end of Appendix~C).\footnote{This arrangement
is similar to the design of CPU chips which have not
only access to their own register memory ($\sim$ the equations known
within {\sc Crack}) and access to the hard disk ($\sim$ the
possibility to call the `probing' module) but which also have
access to cache memory ($\sim$ the buffer file).}

{\em Some more miscellaneous comments.}

As shown in Appendix~C %\ref{probing}
the generated equations often take
the form of products set equal to zero. This leads to many case
distinctions of factors being either zero or non-zero and consequently
other factors being zero. The depth of sub-case levels sometimes
reaches 20. Because buffer files are only valid for the case in
which they were generated (because they make use of case-dependent
known substitutions $q_{{\cal D}}=h_{{\cal D}}(q'_{{\cal D}})$ and case-dependent inequalities) and because of these deep
levels of sub-cases, the number of buffer files easily reaches 100,000
and more (e.g.\ too many to be deleted with the simple UNIX command
{\tt rm *}). Therefore the case label is encoded in the buffer file
name allowing to delete buffer files automatically when the case in
which the file was created and all its sub-cases are solved.

There is much room for experimenting with the place of the two new modules
within the priority list\footnote{apart from the necessity to give the
`reading from the buffer' module a higher priority than the
`creating of a buffer' module in order to try reading and emptying a buffer
file first (if available and not already read completely) and to
create a new buffer file only if none is available
or if the available one is already completely read in}.
On one hand one wants to read and create
early, so that the ongoing computation has many equations to choose
from when looking for the most suitable substitutions, shortenings, ... . The
problem is that these equations are all generated with the same
limited information on relations between $q_{{\cal D}}$, and thus they
have a high redundancy.  Also, dealing with many long equations does slow
down {\sc Crack}. On the other hand, giving the branching of the
computation into sub-cases a higher priority generates an exponential
growth of sub- and sub-sub-cases which drastically increases the
number of buffer files to be generated
because they are only valid for the case they
were generated for or for its sub-cases.

With each investigated case, say $q_5=0$, the other case $q_5\neq 0$
generates inequalities which the package {\sc Crack} collects, updates
and makes heavily use of to avoid further case distinctions as far as
possible, and in this computation also to drop factors of the $\hat{P}_{mi}$
as mentioned in Appendix~D. %\ref{cache}.

The individual cases can either be investigated serially or in
parallel.

If two solutions of two different cases, for example, the solutions
for $q_5=0$ and $q_5\neq 0$ can be merged into one analytic form, if
necessary by a re-parametrization, then this is achieved by one of the
modules of {\sc Crack} (\cite{merge}).

\section*{Appendix E: The Computation} \label{computation} %=============

The computation was not performed in a single run. It fact due to
the big workload it extended over
several months. Simple subcases were solved initially whereas harder ones were
completed only after the probing technique from Appendix~C and its
automatic interplay with the package {\sc Crack} were developed. Even
then it took some time to fine-tune parameters and put the new modules
in the right place within the priority list of procedures in {\sc Crack}.
Finally, it was nearly possible to do the computation fully
automatically, only a few times the proper case distinctions had to be
initiated manually at the right time to be able to complete the computation. If
one would add up purely the necessary computation time without runs
following a poor manual choice of case distinctions leading to
computations which generated too large systems and which could not be
completed then this would amount to 2 weeks of CPU time on a 3GHz AMD64 PC.


\begin{thebibliography}{99}

\bibitem{ABS}
V.E.~Adler, A.I.~Bobenko, Yu.B.~Suris, \textit{Classification of
integrable equations on quad-graphs. The consistency approach}.
Comm. Math. Phys., 2003, v.~233, p.~513--543.

\bibitem{ABS2}
V.E.~Adler, A.I.~Bobenko, Yu.B.~Suris,
\textit{Discrete nonlinear hyperbolic equations. Classification of integrable cases}.
Preprint at \texttt{arXiv:nlin.SI/0705.1663}.

\bibitem{BS}
A.I.~Bobenko, Yu.B.~Suris, \textit{Discrete differential geometry.
Consistency as integrability}.
 Preliminary version of a book (2005).  Preprint at \texttt{arXiv:math.DG/0504358}.

\bibitem{ABOP}
A.I.~Bobenko, Yu.B.~Suris, \textit{On organizing principles of
Discrete Differential Geometry. Geometry of spheres}. Russian Math.
Surveys, 2007, v.~62, $No$~1, p.~1--43.
 Also \texttt{arXiv:math/0608291}.

\bibitem{GT}
E.I.~Ganzha, S.P.~Tsarev, \textit{On superposition of the auto-Baecklund
transformations for (2+1)-dimensional integrable systems}.
Russian Math. Surveys, 1996, v. 51, $No$~6, p. 1200--1202.
See also \texttt{arXiv:solv-int/9606003}.

\bibitem{KSch}
B.G.~Konopelchenko, W.K.~Schief, \textit{ Reciprocal figures,
graphical statics and inversive geometry of the Schwarzian BKP
hierarchy}.
Stud. Appl. Math., 2002, v.~109, p.~89--124.
Also preprint \texttt{arXiv:nlin.SI/0107001}.

\bibitem{NSch}
J.J.C.~Nimmo, W.K.~Schief,
\textit{An integrable discretization of a $2+1$-dimensional sine-gordon
equation.}
\newblock {Stud. Appl. Math.}, 1998, v.~100, p.~295--309.

\bibitem{Veselov}
V.G. Papageorgiou, A.G. Tongas,  A.P. Veselov,
\textit{{Y}ang-{B}axter maps and symmetries of integrable equations on
quad-graphs.}
\newblock {J. Math. Phys.}, 2006, v.~47(8), $No$~083502, 16 pp.

\bibitem{Pot}
Y. Liu, H. Pottmann, J. Wallner, Y. Yang, W. Wang, \textit{Geometric
modelling with conical meshes and developpable surfaces.}
ACM Trans. Graphics. 25(3), 681-689 (2006), Proc. SIGGRAPH 2006.
See also \\ \texttt{http://dmg.tuwien.ac.at/pottmann/}

\bibitem{FORM} J.A.M. Vermaseren, \textit{New features of FORM}.
arXiv:math-ph/0010025, a complete distribution can be downloaded
from   \texttt{http://www.nikhef.nl/\~{}form/}.

\bibitem{CrackOver}  T. Wolf,
\textit{Applications of CRACK in the Classification of Integrable Systems}, in:
CRM Proceedings and Lecture Notes, v.~37 (2004) pp.~283--300. \\
Also \texttt{arXiv: nlin.SI/0301032}.

\bibitem{CrackOnline}
T. Wolf, \textit{An Online Tutorial for the package} {\sc Crack}. \\
{\tt http://lie.math.brocku.ca/crack/demo}

\bibitem{merge}
T.~Wolf.
\newblock Merging solutions of polynomial algebraic systems.
\newblock Preprint, \newline
  \texttt{http://lie.math.brocku.ca/twolf/papers/merge-sig.ps}, 2003.



\end{thebibliography}
\end{document}